# Single step synthesis of size-controlled carbon quantum dots using electrochemical etching of graphite

Ashish Gaurav, Deepali Sinha, Rakesh K. Prasad, Dilip K. Singh*

Birla Institute of Technology Mesra, Ranchi, India-835215

Email: dilipsinghnano1@gmail.com

## Abstract

Carbon Quantum dots (CQD's) are nanoscale $sp^2$ hybridized carbon particles. In this work, we present a simple one-step synthesis of CQDs from the electrochemical shredding method and technique to control its size during its growth process. A graphite rod extracted from commercially available pencil batteries was used as electrode: source of carbon. CQDs of varying sizes were synthesized through controlled current to the solution for etching and Sodium dodecyl sulfate (SDS) as capping agent during growth. CQD's of controlled sizes as formed can be used as fluorescent marker for bio-imaging and sensing platform for wide range of applications.

## Introduction

Carbon Quantum dots (CQDøs) are innovative class of carbon nanomaterial with various promising properties. With quasi-spherical shape and size below 10 nm they form traditional semiconductor quantum dot due to integration of optical properties with electronic properties of the carbon. In contrast to semiconductor quantum dots which are cytotoxic towards living body cells, CQD's have the advantage of chemical inertness and highly reduced cytotoxicity **[1]**. CQD's have a carbogenic core with different functional groups on its surface which is basically composed of nano-crystalline and amorphous core with a high percentage of $sp^2$ hybridized carbon and the lattice sites are generally occupied by graphitic carbon, along with the presence of oxygen functional groups at the termination of CQD particle results for higher stability in aqueous solution **[2]**. Generally, CQD's due to attracting features like low chemical toxicity, high chemical inertness along with high stability makes it a matter of great attentiveness. Researchers have been exploring variety of quantum dots to find a suitable alternative of organic dye molecules which have limited photo-stability and suffer from photo-blinking behavior. In comparison to traditional dyes, CQD's are much brighter and more photo-stable and are promising in multiplexing application due to their narrow fluorescence emission spectra.

The properties of CQD's largely depends on the size, morphology, type and doping amount. Due to these novel properties CQD's have been widely used in various fields like bio-medicine, energy, sensing etc. As compared to GQD's (Graphene quantum dots), CQDs have poorer crystallinity and have lesser number of $sp^2$ hybridized carbon resulting into higher density of defects **[3, 4]**. The presence of immense amount of oxygen containing functional group on the surface are very useful for surface modification via covalent bonding **[5, 6]**. Surface modification is a vital phenomenon which efficiently modifies the physical and optical properties of quantum dots highly desirable for sensing applications as it enables selective sensing. Fluorescence emission wavelength of various CQDs or GQDs is very much sensitive to size, different surface defects, surface states and the extent of degree of conjugation. Due to their intrinsic fluorescence properties, CQDs and GQDs can be used as a sensor which can be further divided into various categories like luminescence sensor, electrochemical sensors, optical bio-sensors and electrochemical bio-sensors etc **[7]**. Till date, CQDs have been widely used in the detection of heavy metal ions or pesticides **[8, 9]** and changing their fluorescence properties also gave a promising result in the detection of inorganic molecules **[10]**. The high quantum yield plays an important role in the action of CQD or GQD as a fluorescent probe or sensor.

GQDøs can be synthesized by oxidative cleavage **[11]** solvothermal method **[12]**, electrochemical oxidation **[13]**, micro-wave or ultrasonic assisted techniques **[14],** carbonization from tiny molecules and Chemical vapor deposition (CVD) etc. Alternatively, using top down approach GQDøs can be synthesized using graphene, graphitic rods, carbon nanotubes etc and are obtained by cutting them down to nanoscale using diverse techniques like laser ablation, hydrothermal process or electrochemical process etc. In case of bottom up approach, due to variety of parameters being involved synthesis of monodisperse GQDøs is difficult **[15].** In 2019, Fu et al. adopted an electrochemical exfoliation method of graphitic rod considering them as electrodes in a chemical solution containing NaOH and Semicabazide in a fixed proportion to obtain N-doped GQD **[16]**. Recently using non-aqueous media of propylene carbonate with $LiClO_4$ graphene quantum dot was synthesized using electrochemical approach **[17]**. In this article, we have used single step electrochemical etching process for the synthesis of GQDøs using graphitic rod.

**Experimental Details**

Synthesis of Carbon Quantum Dots:

In our experiment, we used top-down approach to get the desired CQDøs. The CQDøs are prepared by electrochemical synthesis process using NaOH schematically as shown in Fig. 1(a). The precursors used were Ethanol and NaOH. 99.5 ml of Ethanol was taken in a 500 ml beaker. 0.5 ml of distilled water was supplemental to it which was followed by the addition of 0.25 g - 0.33 g of NaOH palette. The solution was then set to magnetic stirring at 750 rpm till NaOH dissolves completely. 50 mL of so obtained solution was used as the electrolyte for the synthesis. Graphite rods were extracted from commercial batteries (Eveready AA pencil cells 1.5 volts). These graphite rods were used as both anode and cathode for the electro-chemical process. Preferably, 70% of the graphite rod was dipped in the electrolyte during the process. A current of 30-45 mA was supplied continuously for 2-3 hours. The solution obtained after the electro-chemical process was stored in a glass vial as shown in Fig. 1(b). After the 8-9 hours, the solution changes color from colorless to light yellow as shown in Fig. 1(c).

The rest 50 mL of the solution prepared to be used as the electrolyte was taken in a beaker followed by the addition of $1 \times 10^{-3}$ mole of Sodium dodecyl sulphate (Sigma Aldrich, RG). The solution was stirred for 10 minutes. This solution was used as a limiting agent for the growth control of CQDøs by the mechanism of capping of the CQDs solution. Etched CQD solution was divided into three equal parts. Out of which 2 parts were capped using at 4 hours, 8 hours and 14 hours of growth referred as CQD4H and CQD8H and CQD14H. To seize the further growth of etched CQD solution, 4 ml of SDS solution was added to 6 ml of etched CQD solution.

**Mechanism of Growth:**

During this one step electrochemical shedding process, two main process occurs which includes cutting and oxidation at the defect site or at the surface of the Graphite rod dipped in the solution which acts as an anode and cathode **[16]**. Applied current breaks the water molecule formed as a product of the electrolyte solution into $H^+$ and $OH^-$ and the anion ($OH^-$) ion insinuate in between the graphitic layers of graphite rod and as a result of which the $OH^-$ ion get oxidized and produces oxygen which creates a pressure in between the layers and they occupy the Van-der Waals gap. As an outcome there is peeling of the upper layer of the graphitic layer and the cleavage of C-C bond gives out the CQD's in the solution **[18].**

**Results and discussion**

Fig.2 (a) and (b) shows the FESEM image of CQD's formed using current density of 35 mA at two magnifications 200 k× and 400 k×. It shows monodispersed carbon particles of size of < 10 nm. Fig. 3 shows the UV-Vis spectra of CQDs produced at varying current density 20 mA, 35 mA and 46 mA. The absorption peak monotonically shifts to higher wavelength with increasing the current density. The absorption band-edge is observed at 2.65 eV, 2.25 eV and 2.18 eV for CQDs formed with 20 mA, 35 mA and 46 mA respectively as estimated using tauc plot shown in Fig. 3(b). It is evident from varying absorption peak that size of the CQDs formed depends upon the current density used for etching. This indicates that with increasing applied current density graphitic sheet breaks into smaller size CQDs as evident from blue shift of absorption peaks with increasing current density.

Raman spectra of synthesized CQD's is shown in Fig. 4(a). The experimentally obtained curve was fitted using Lorentzian line shape as shown in Fig. 4(b) and the line profile is summarized in table-1. The fitted curve shows prominent peaks (FWHM) at 1329 ($\Delta\omega=95$) $cm^{-1}$ and 1380 ($\Delta\omega=167$) $cm^{-1}$ (referred as D-bands), 1551 ($\Delta\omega=431$) $cm^{-1}$ (G-band) and 1863 ($\Delta\omega=281$) $cm^{-1}$ (M-band) respectively. In addition to it, a weak shoulder is observed at ~ 1170 cm$^{-1}$ with (Δω=164) cm$^{-1}$ referred as defect peaks. In literature such peaks have been accounted to pentagon-heptagon pairs **[19]**. D-band originates from A$_{1g}$ ósymmetric mode activated by loss of translational symmetry due to defects. It was first attributed to an A$_{1g}$ breathing mode at K, activated by the relaxation of the Raman fundamental selection rule q = 0 **[20]**. The double resonance process at k-point in graphite gives rise to D-band **[21]**. Peak at 1331.8 cm$^{-1}$ is assigned to triply degenerate (F$_{2g}$) zone center phonon mode of diamond lattice (arising from O$^7_h$ space group) **[22]**, while peak at 1351.5 cm$^{-1}$ with relatively larger peak width is assigned to dis-ordered *sp$^3$* bonded carbon structure/ amorphous carbon **[23]**. Peak at 1351.5 cm$^{-1}$ arises from lattice distortion induced breakdown of momentum conservation requirement, which activates phonons at the interior K point of the Brillouin zone (q≠0) **[24]**. Peak at ~1578.8 cm$^{-1}$ is assigned to the graphitic G-mode of carbon arising from intra-layer doubly degenerate E$_{2g}$ mode of *sp$^2$* bonded graphitic carbon structure. The peak width of D-line remain nearly same (~95 cm$^{-1}$ for 1329 cm$^{-1}$ peak and 166 cm$^{-1}$ for 1380 cm$^{-1}$ peak) indicating no change in the crystallinity of nanoparticles formed at different current density. Also the I$_D$/I$_G$ ratio of these peaks remains same. Large peak-width of observed peaks indicates amorphous nature of the formed nanoparticles. Sample synthesized using 35 mA current shows much stronger Raman intensity for both D-band and G-band as compared to CQDøs grown using higher current densities.

Fig. 5 shows PL spectra of synthesized CQDs. The PL spectra shifts from red to violet region as the size of the CQDøs increases. HOMO-LUMO gap of CQDøs depends upon the size of the quantum dots and it gives rise to strong emission intensity **[25]**. The PL spectra of the CQDs shows that the emission wavelength of CQDøs shifts towards ultraviolet on increasing the current density supplied to the graphite electrode during the electrochemical process. This shows that the size of the CQDs decreases with increase in the current density. Also, it is observed that the PL intensity increases with increase of the supplied current density. Small size CQDøs emits strongly than the larger size dots. The observed PL peak position is in the opposite order to that of observed absorption spectra. Particle synthesized with 20 mA shows highest energy of absorption (468 nm, i.e 2.65 eV) and minimum energy of emission (543 nm, i.e 2.28 eV). While CQDs synthesized using 46 mA shows absorption at lowest energy 2.18 eV and emission peak at 2.43 eV, at highest energy among the three used current densities.

Further, the results of attempt to control and fine tune the size of CQDs using capping during growth is shown in Fig. 6. Post electrochemical etching of graphite, formed CQD capped after 4 hours, 8 hours and 14 hours of nucleation process is shown in Fig. 6(a). CQD14H refers to the CQDs which were allowed to grow for larger duration i.e completely as indicated by further no color change after synthesis with current density of 35 mA. The UV-Vis spectra of the capped and uncapped are shown in figure 6 (b). The uncapped CQDs show absorption/ excitation peak at 464 nm while that of the capped CQDs is at 424 nm. This states that the size of the particle which was capped is smaller as compared to the nanoparticles present in the solution which was allowed to grow. Therefore, the addition of SDS seizes further growth by forming micelles. The micro-emulsion formed upon addition of SDS controlled the growth by capping the CQDøs from all the direction resulting into smaller size particles. Figure 6(c) shows the tauc plot of the capped and uncapped CQDs. Particle capped after 04 hours shows higher band-edge of absorption (2.4 eV) than the uncapped particle grown for 14 hours (2.1 eV). Absorption at higher energy indicates smaller particle size due to quantum confinement effects, confirming the possibility to further fine tune the size of the nanoparticles with growth time. Fig. 6(d) shows the PL spectra of capped and uncapped quantum dot

CQD4H and CQD14H respectively. The smaller size CQD (CQD4H) shows much stronger emission at 515 nm than the CQD14H (536 nm).

Fig. 7 shows the FTIR spectra of capped and uncapped CQDs. The capped and uncapped CQDs (CQD4H and CQD14H shows similar peaks, indicating presence of similar functional groups on the surface. Peaks are observed at 3433.14 $cm^{-1}$ (óOH vibrations), at 1384.88 $cm^{-1}$ (óCO vibrations) and at 644.22 $cm^{-1}$ (C-O-C vibrations), 1643 $cm^{-1}$ (C=O) and at 2927 $cm^{-1}$ (C-H bond). Due to presence hydroxyl and carboxylic groups on the surface as confirmed using FT-IR, the synthesized CQDøs are water dispersible .After addition of SDS to the solution of CQDs during growth, no new vibrational peak were observed in FTIR. This indicates that the SDS solution does not react with CQDs and thus no new bonds are formed. There is only a physical interaction between the CQDs and SDS.

**Conclusion**

Size controlled synthesis of CQDs was successfully achieved through single step electro-chemical etching of graphite electrodes extracted from pencil batteries. Current density used for etching graphitic electrode is observed to influence the size of the CQDs formed. Further, the size of the CQDs was fine-tuned through use of SDS as surfactant during nucleation of etched carbon atoms towards formation of quantum dot. The optical absorption and emission spectra of synthesized CQDs shows monotonic variation with increasing size. These optically active CQDs would be highly useful as imaging contrast material for selective bio-imaging, as optical absorber and in photovoltaics.

**Acknowledgement**

We are thankful to TEQUIP-III for grants CRS ID -1-5736483014, CRS ID-1-5743255881 and DST Govt of India for grant IFA13-PH65.

**Table-1 Fitted Raman peak profile.**

| Sample / | | Peak-1 | Peak-2 | Peak-3 | Peak-4 | Peak-5 |
|---|---|---|---|---|---|---|
| 15 mA | Centre | 1170.53 | 1329.74 | 1380.43 | 1551.61 | 1863.34 |
|  | FWHM | 164.319 | 95.4019 | 166.67 | 430.645 | 281.054 |
|  | Ampl. | 7108.54 | 27382.5 | 33574.5 | 34960.9 | 18218.1 |
| 20 mA | Centre | 1170.69  | 1329.61  | 1380.24  | 1550.7   | 1864.45  |
|       | FWHM   | 166.582  | 94.3568  | 165.558  | 437.641  | 273.021  |
|       | Ampl.  | 6767.32  | 26506.5  | 32431.4  | 33475.4  | 16586.9  |
| 35 mA | Centre | 1171.53  | 1330.39  | 1380.92  | 1552.42  | 1862.79  |
|       | FWHM   | 164.446  | 95.7706  | 168.831  | 427.813  | 272.042  |
|       | Ampl.  | 11621.7  | 44203.5  | 53890.8  | 54475.2  | 26035.5  |
| 46 mA | Centre | 1170.81  | 1330.2   | 1381.35  | 1554.72  | 1864.21  |
|       | FWHM   | 162.851  | 95.1995  | 168.6    | 428.62   | 272.676  |
|       | Ampl.  | 8008.94  | 31725.1  | 38725.7  | 39634.9  | 20568.8  |
|       |        | Defects  | D-band   |          | G-band   | M-band   |

**Table-2 Fitted PL peak profile.**

| Sample / Peak |        | Peak-1  | Peak-2  | Peak-3  | Peak-4   |
|---------------|--------|---------|---------|---------|----------|
| 20 mA         | Centre |         |         | 542.429 | 613.658  |
|               | FWHM   |         |         | 93.9251 | 226.333  |
|               | Ampl.  |         |         | 4.25771 | 0.535098 |
| 35 mA         | Centre |         |         | 530.468 | 591.565  |
|               | FWHM   |         |         | 85.285  | 96.2218  |
|               | Ampl.  |         |         | 29.0367 | 5.65797  |
| 46 mA         | Centre | 472.954 | 511.66  | 545.82  |          |
|               | FWHM   | 42.0696 | 86.013  | 122.191 |          |
|               | Ampl.  | 15.2455 | 91.1893 | 26.3428 |          |
| 4H-capped     | Centre | 471.336 |         | 530.827 | 592.647  |
|               | FWHM   | 13.5169 |         | 84.7465 | 95.1289  |
|               | Ampl.  | 0.94037 |         | 29.2071 | 5.60141  |
| 14H-capped    | Centre | 474.959 | 514.121 |         | 581.436  |
|               | FWHM   | 36.7888 | 93.397  |         | 103.836  |
|               | Ampl.  | 2.97016 | 50.4003 |         | 5.75903  |

**Figure Captions**

Figure 1: (a) Schematic diagram of the experimental setup used for synthesis of CQDs. (b) CQDs sample at initial stage of electrochemical synthesis and (c) fully grown sample after aging.

Figure 2 shows the FESEM image of the CQD's formed at current density of 35 mA and at magnifications (a) 200 k× and (b) 400 k×.

Figure 3(a) UV-Vis spectra of CQDs produced at different current density with intensity 20 mA, 35 mA and 46 mA. (b) Corresponding Tauc plot shows band-gap of 2.65 eV, 2.25 eV and 2.18 eV respectively for samples grown with 20 mA, 35 mA and 46 mA respectively.

Figure 4. (a) The Raman Spectroscopy of CQDs synthesized using electrochemical etching of pencil rods with varying current density. (b) Fitted spectra shows peaks at 1172 cm$^{-1}$ (defects), 1330 cm$^{-1}$ & 1381 cm$^{-1}$ (D-band), 1552 cm$^{-1}$ (G-band) and 1812 cm$^{-1}$ (M-band).

Figure 5. Photoluminescence of carbon quantum dots produced using current densities 20 mA, 35 mA and 46 mA.

Figure 6 (a) Photograph of CQDs capped with SDS during growth after 4 hours (CQD4H), 8 hours (CQD8H) and after 12 hours (CQD12H). (b) UV-Vis spectra of uncapped and capped CQDs. (c) corresponding tauc plot shows band gap of 2.1 eV and 2.42 eV respectively for the capped and uncapped CQDs . (d) PL spectra of CQD4H and CQD12H. As the particle grows emission shifts towards red wavelength and emission intensity monotonically decreases.

Figure 7. FTIR spectra of SDS capped and uncapped CQDs indicate no change in the chemical functional group upon addition of surfactant.

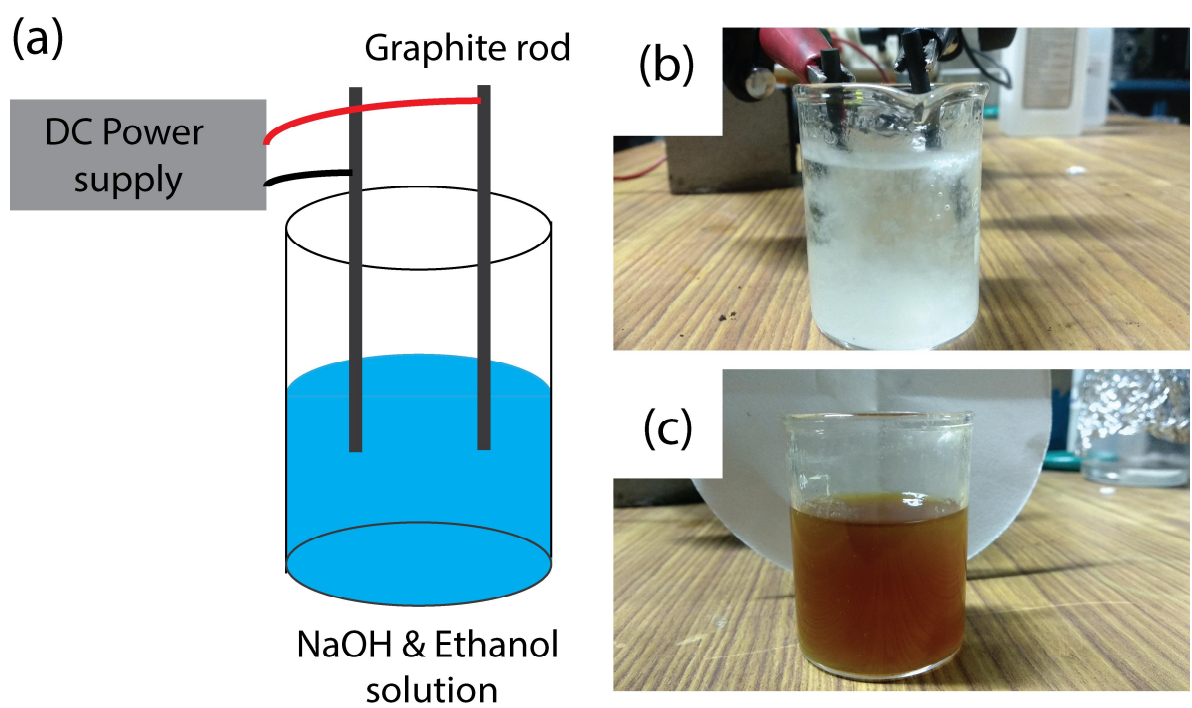

**Fig. 1**

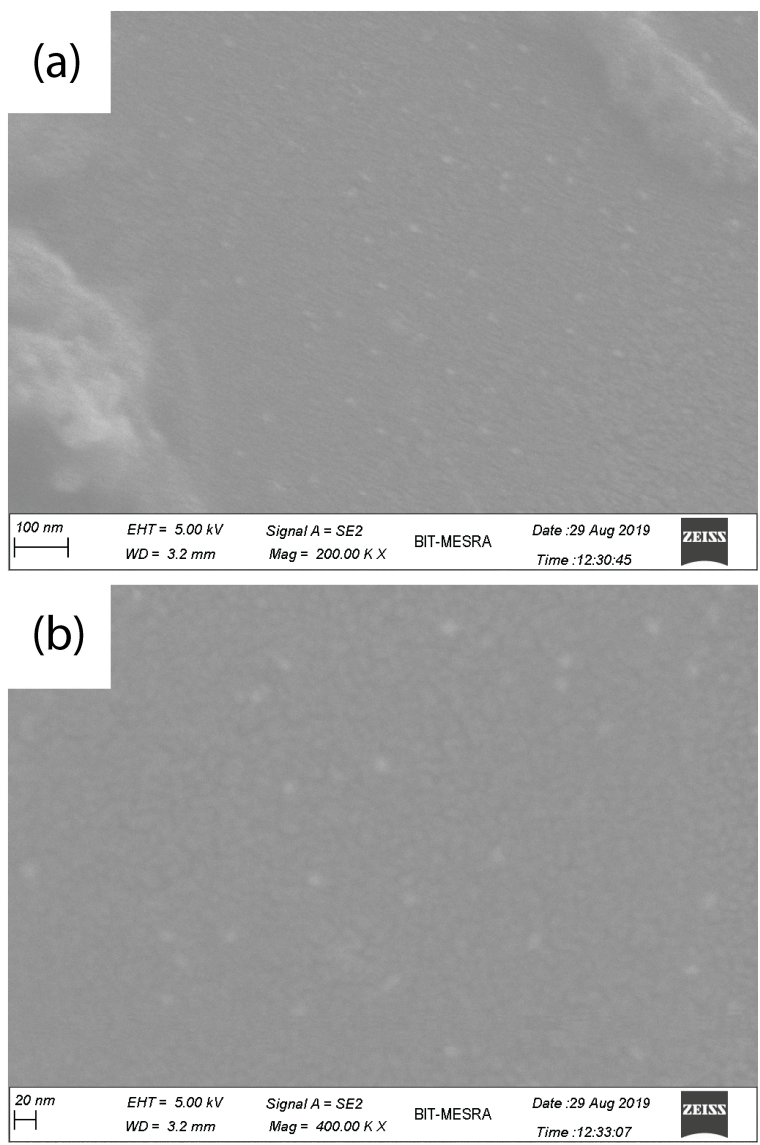

**Fig. 2**

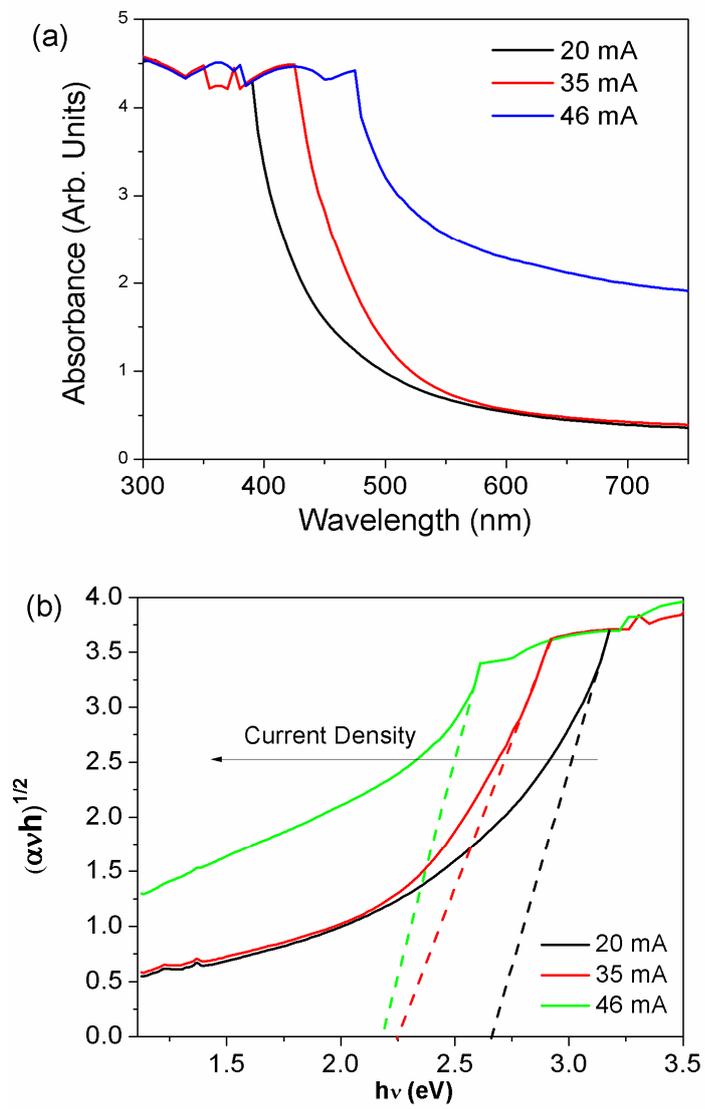

**Fig. 3**

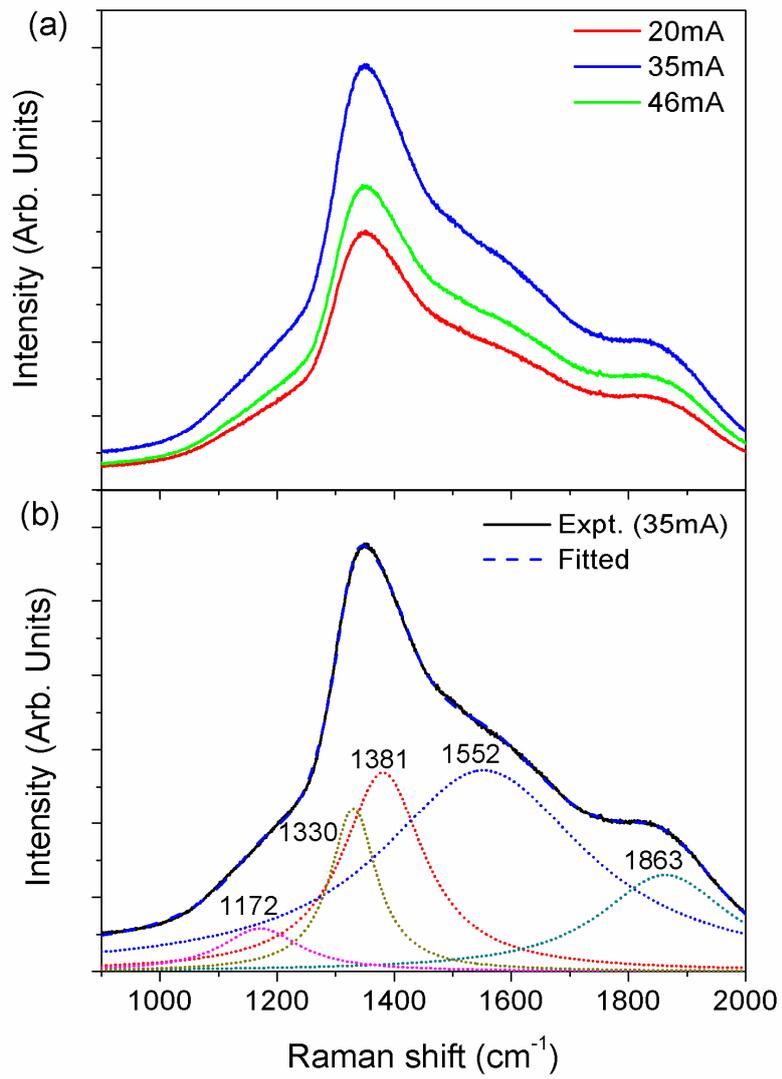

**Fig. 4**

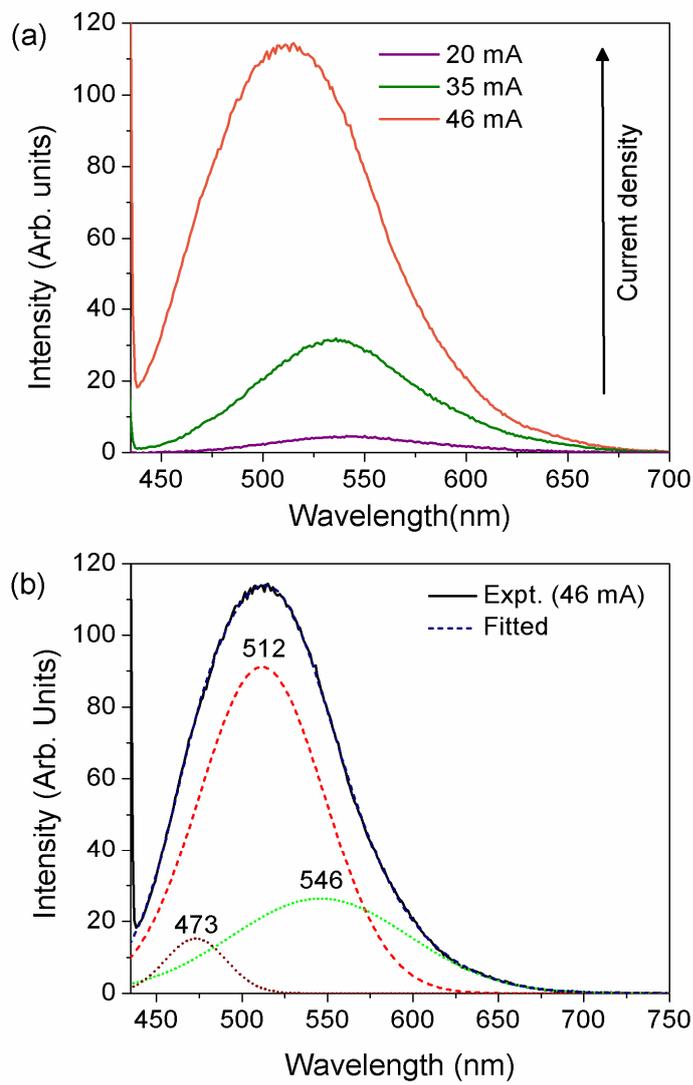

**Fig. 5**

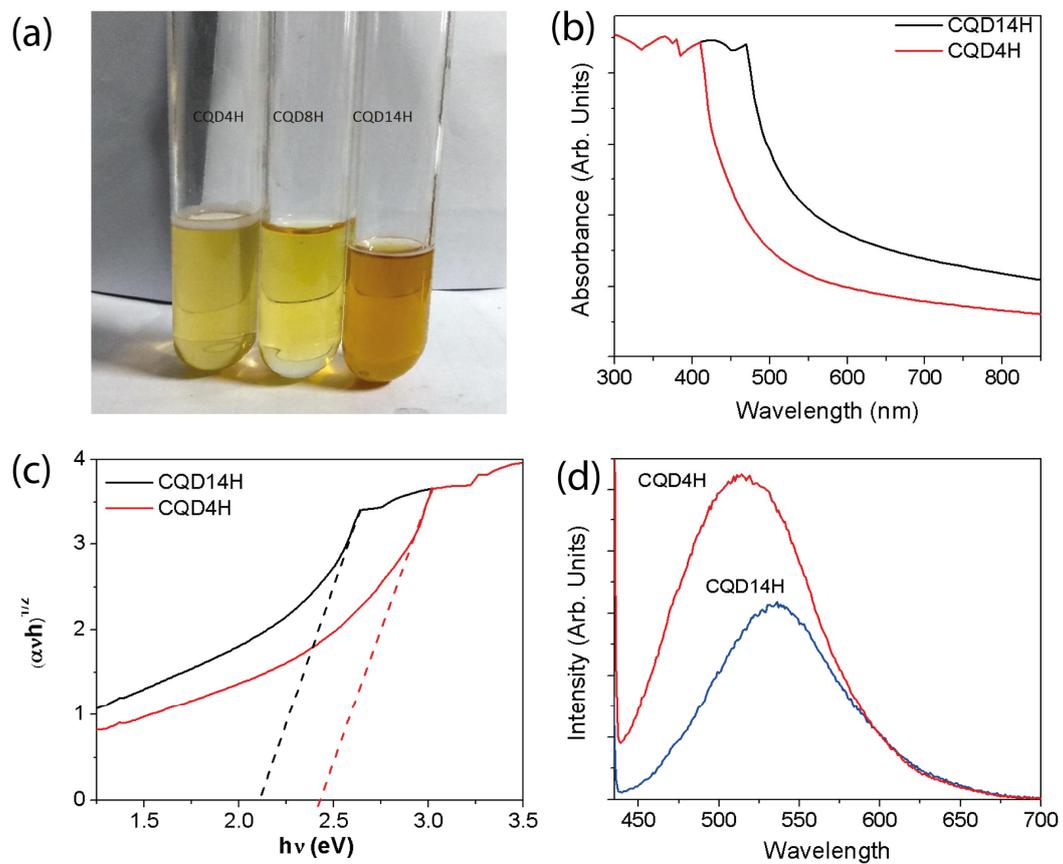

**Fig. 6**

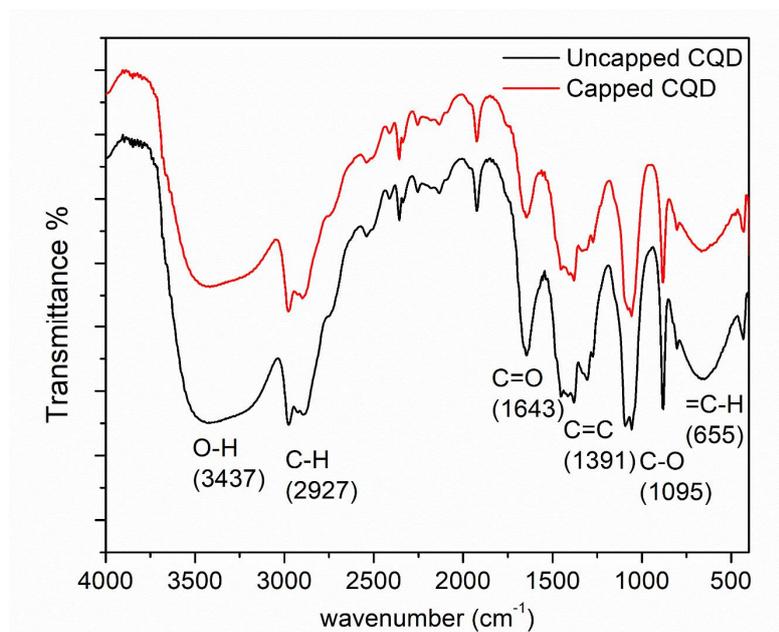

**Fig. 7**